\documentclass[aps]{revtex4}
\usepackage{eurosym}
\usepackage{amsfonts}
\usepackage{amsmath}
\usepackage{amssymb,epsf}
\usepackage{color}
\usepackage{xcolor}
\usepackage{graphicx}
\usepackage{epstopdf}
\usepackage{float}
\usepackage{caption}
\usepackage{subfig}

\begin{document}

\title{Phantom BTZ black holes}
\author{B. Eslam Panah$^{1}$\footnote{%
email address: eslampanah@umz.ac.ir}, and M. E. Rodrigues$^{2,3}$ \footnote{email address: esialg@gmail.com}}
\affiliation{$^{1}$ Department of Theoretical Physics, Faculty of Science, University of
Mazandaran, P. O. Box 47416-95447, Babolsar, Iran\\
$^{2}$ Faculdade de Ci\^{e}ncias Exatas e Tecnologia, Universidade Federal do Par\'{a}\\
Campus Universit\'{a}rio de Abaetetuba, 68440-000, Abaetetuba, Par\'{a}, Brazil\\
$^{3}$ Faculdade de F\'{\i}sica, Programa de P\'{o}s-Gradua\c{c}\~{a}o em F\'{\i}sica, Brazil}

\begin{abstract}
Motivated by the impact of the phantom field (or anti-Maxwell field) on the structure of three-dimensional black holes in the presence of the cosmological constant, we present the first extraction of solutions for the phantom BTZ (A)dS black hole. In this study, we analyze the effect of the phantom field on the horizon structure. Furthermore, we compare the BTZ black holes in the presence of both the phantom and Maxwell fields. Additionally, we calculate the conserved and thermodynamic quantities of the phantom BTZ black holes, demonstrating their compliance with the first law of thermodynamics. Subsequently, we assess the effects of the electrical charge and the cosmological constant on the local stability in the canonical ensemble by considering these fields with respect to the heat capacity. We then investigate the global stability area of the BTZ black holes with phantom and Maxwell fields within the grand canonical ensemble using Gibbs free energy. In this analysis, we evaluate the influence of the electrical charge and the cosmological constant on this area.
\end{abstract}

\maketitle

\section{introduction}

Black holes are celestial objects characterized by strong gravitational
fields and intricate structures. They play an important role in the
investigation of theories of gravity. According to the theory of general
relativity (GR), a black hole is defined as a region in spacetime
encompassed by an event horizon. Hawking further expanded on this concept by
demonstrating that the area of an event horizon cannot decrease over time 
\cite{Hawking1972}. In this regard, Bekenstein also formulated the concept
of black hole entropy \cite{Bekenstein1972,Bekenstein1973,Bekenstein1974}.
It has also been established that there is a correlation between surface
gravity and temperature, indicating that surface gravity can be utilized to
determine the temperature of a black hole \cite{Bardeen1973}. Hawking
proposed a mechanism through which black holes can emit thermal radiation,
which is known as Hawking temperature, proportional to their surface
gravity. This radiation is caused by quantum fluctuations near the event
horizon, which allow for creating virtual particle pairs and observing
positive energy flux in distant regions \cite{Hawking1974}. On the other
hand, recent observations have provided compelling evidence supporting the
existence of black holes in our universe \cite%
{Abbott2016,Akiyama2019a,Akiyama2019b,Abbott2020}. So, Black holes were
regarded as thermodynamic systems, wherein the temperature and entropy are
denoted by the surface gravity and area of the black hole horizon,
respectively. The study focused on the phase transition of a black hole as a
thermodynamic system. Specifically, it was demonstrated that a Hawking-Page
phase transition exists between the pure radiation phase and the stable
large Schwarzschild black hole phase in AdS space \cite%
{Hawking1983,Witten1998}. Additionally, a first-order phase transition,
resembling the liquid-gas system, was observed between the small and large
black hole phases for charged AdS black holes \cite{Kubiznak2012}. Numerous
studies have been conducted to explore this phenomenon, considering
different types of black holes \cite%
{PV1,PV2,PV3,PV4,PV5,PV6,PV7,PV8,PV9,PV10,PV11,PV12,PV13,PV14}.

The first three-dimensional black hole solutions were introduced by Banados, Teitelboim, and Zanelli which are known as BTZ black holes \cite{BTZ}. The significance of BTZ black holes lies in their ability to provide an elegant framework for understanding the interactions of lower-dimensional gravitational systems \cite{Witten2007}. They have also been utilized to establish a link with string theory \cite{Witten1998}, study different thermal properties of black holes \cite{Cadoni2009,Sarkar2006}, and see Refs. \cite{BTZ}, for more details. Since then, different theories of gravity to investigate various three-dimensional black hole solutions have been explored in much literature  \cite{BTZ1,BTZ2,BTZ3,BTZ4,BTZ5,BTZ6,BTZ7,BTZ8,BTZ9,BTZ10,BTZ11,BTZ12,BTZ13,BTZ14,BTZ15,BTZ16,BTZ17,BTZ18,BTZ19,BTZ20,BTZ21,BTZ22,BTZ23,BTZ24,BTZ25,BTZ26,BTZ27,BTZ28,BTZ29}. The study of black holes in three dimensions can help us understand the phenomena that occur in black hole physics in four dimensions. We have three main motivations for doing a study in a lower dimension: 1- While gravity is not perturbatively renormalizable in four spacetime dimensions, it is in three dimensions. Notably, three-dimensional gravity possesses black hole solutions and is exactly solvable, paving the way for the exploration of quantum black holes. This makes three-dimensional gravity an intriguing system in its own right; 2- The AdS/CFT correspondence establishes a connection between conformal field theories (CFT) in two dimensions and gravity in three dimensions. CFTs play a significant role in condensed matter physics, and studying 3D gravity can provide valuable insights into these theories; 3- Gravity in three dimensions is simpler to handle compared to gravity in four dimensions, making it an effective toy model for understanding four-dimensional gravity. In quantum gravity, models in $2+1$ dimensions are of great importance for understanding new phenomena \cite{3d,3d2,3d3,3d4,3d5,3d6,3d7}. According to the importance of BTZ black holes, we study the phantom BTZ black holes.

Observational evidence indicates that our Universe is accelerating, and the
concept of dark energy was proposed to explain this accelerated expansion 
\cite{Dark1,Dark2,Dark3,Dark4}. Considering that our Universe is filled with a perfect fluid of pressure $p$ and density $\rho$, an important property of this fluid is that by the equation of state (EoS) $p=\omega \rho$, the fluid has $\omega\sim -1$ today. So the fluid has exotic characteristics with negative pressure and attractive gravity. By applying the Seven-Year Wilkinson Microwave Anisotropy Probe (WMAP) observations data, the latest distance measurements from the BAO in the distribution of galaxies, and the Hubble constant measurement, for a flat universe, the current value of a constant EoS for dark energy has been estimated as $\omega=-1.10\pm 0.14$ ($68\%$ confidence level) \cite{wmap}. The model most consistent with the observations is $\Lambda$CDM \cite{LCDM}, where we can have a phantom phase with $\omega<-1$, asymptotically de Sitter phantom universe \cite{DE}. This phantom phase can be modeled using a scalar field with a kinetic term in the action, opposite to the conventional canonical one. In this case, the pressure of the field is negative \cite{PF,PF2}. One specific type of dark energy, called the phantom field, exists in the Einstein-Maxwell-dilaton system \cite{EMD}  and is studied in relation to ghost branes in string theory \cite{String}. It should be noted that the phantom field may give rise to quantum instabilities \cite{Dark1,Dark2}, presenting a challenge to the theory. However, in Refs. \cite{Quantumph1,Quantumph2} argued that these instabilities can be avoided. Therefore, the phantom field has emerged as a potential candidate for the dark energy model from a theoretical perspective. Given the significance of the phantom field, and particularly its effects on the background of the black hole spacetime, this investigation is important because dark energy will inevitably influence black holes in the Universe. on the other hand, a new black hole solution known as the anti-Reissner-Nordstrom-(A)dS solution or phantom Reissner-Nordstrom-(A)dS black hole solution \cite{PhantomBH} was extracted in the Einstein-anti-Maxwell theory in the presence of a cosmological constant. These black holes possess a phantom charge, making them fundamentally distinct from the usual Reissner-Nordstrom-(A)dS black holes. Therefore, studying the characteristics of these black holes is of great interest. Historically, Einstein and Rosen proposed the concept of changing $q\rightarrow -q$ in 1935 \cite{EinsteinR}. They suggested adding a purely imaginary charge, to the Reissner-Nordstrom solution to describe what is known as the quasicharged bridge. The literature extensively discusses
phantom black hole solutions in general relativity and other modified
theories of gravity \cite{PhBH1,PhBH2,PhBH3,PhBH4,PhBH5}. So, in this study,
our objective is to investigate the phantom BTZ black hole. To achieve this,
we will first extract the exact solution of the phantom BTZ black hole.
Subsequently, we will analyze other properties, including conserved and
thermodynamic quantities, heat capacity, and phase transition.

\section{Black Hole Solution}

The action of this theory in three-dimensional spacetime is given by 
\begin{equation}
\mathcal{I}=\frac{1}{2\kappa ^{2}}\int_{\partial \mathcal{M}}d^{3}x\sqrt{-g}%
\left[ R-2\Lambda +\eta \mathcal{F}\right] ,  \label{action}
\end{equation}%
where $R$ and $\Lambda $ are, respectively, the Ricci scalar curvature and
the cosmological constant. The third term is the coupling with the Maxwell
field, when $\eta =1$, or a phantom field of spin $1$, when $\eta =-1$.
Also, $\mathcal{F}=$\ $F_{\mu \nu }F^{\mu \nu }$ is the Maxwell invariant.
In addition $F_{\mu \nu }=\partial _{\mu }A_{\nu }-\partial _{\nu }A_{\mu }$
is the electromagnetic tensor field, and $A_{\mu }$ is the gauge potential.
In the above action, $\kappa ^{2}=8\pi G$, and $G$ is the Newtonian
gravitational constant. Hereafter, we consider $G=c=1$. Meanwhile, $%
g=det(g_{\mu \nu })$ is the determinant of metric tensor $g_{\mu \nu }$.

Making the functional variation of the above action (\ref{action}), with
respect to the gauge field $A_{\mu }$ and the gravitational field $g_{\mu
\nu }$, we get the following field equations in the following forms%
\begin{eqnarray}
G_{\mu \nu }+\Lambda g_{\mu \nu } &=&2\eta \left( \frac{1}{4}g_{\mu \nu }%
\mathcal{F}-F_{\mu }^{~\alpha }F_{\nu \alpha }\right) ,  \label{Eq1} \\
&&  \notag \\
\partial _{\mu }\left( \sqrt{-g}F^{\mu \nu }\right) &=&0,  \label{Eq2}
\end{eqnarray}%
where $G_{\mu \nu }$ is the Einstein tensor.

We consider a three-dimensional static spacetime in the following form 
\begin{equation}
ds^{2}=-g(r)dt^{2}+\frac{dr^{2}}{g(r)}+r^{2}d\varphi ^{2},  \label{Metric}
\end{equation}%
where $g(r)$ is the metric function that we have to find it. In order to
obtain phantom black hole solutions, we consider a radial electric field
which its related gauge potential is given 
\begin{equation}
A_{\mu }=h\left( r\right) \delta _{\mu }^{t}=\left( h\left( r\right)
,0,0\right) .  \label{Amunu}
\end{equation}

Using the Maxwell equations (\ref{Eq2}) with the metric (\ref{Metric}), and
the mentioned gauge potential (\ref{Amunu}), we can find the following
differential equation by 
\begin{equation}
rh^{\prime \prime }(r)+h^{\prime }(r)=0,  \label{hh}
\end{equation}%
where the prime and double prime are, respectively, the first and the second
derivatives with respect to $r$. We extract the solution of the equation (%
\ref{hh}) in the following form 
\begin{equation}
h(r)=-q\ln \left( \frac{r}{l}\right) ,  \label{h(r)}
\end{equation}%
where $q$ is a integration constant which is related to the electric charge.
To have logarithmic arguments dimensionless, we introdcue $l$ in which is an
arbitrary constant with length dimension. Considering the equation (\ref%
{h(r)}), the electromagnetic field tensor is given 
\begin{equation}
F_{tr}=\partial _{t}A_{r}-\partial _{r}A_{t}=\frac{q}{r}.  \label{Ftr}
\end{equation}

Substituting (\ref{Ftr}) into the equation (\ref{Eq1}), we get

\begin{eqnarray}
eq_{tt} &=&eq_{rr}=rg^{\prime }(r)+2\Lambda r^{2}-2\eta q^{2},  \label{eq1}
\\
&&  \notag \\
eq_{\varphi \varphi } &=&r^{2}g^{\prime \prime }(r)+2\Lambda r^{2}+2\eta
q^{2},  \label{eq2}
\end{eqnarray}%
where $eq_{tt}$, $eq_{rr}$ and $eq_{\varphi \varphi }$, are components of $%
tt $, $rr$ and $\varphi \varphi $ of field equations (\ref{Eq1}),
respectively.

Using the equations (\ref{eq1}) and (\ref{eq2}), we obtain the metric
function in the following form

\begin{equation}
g(r)=-m_{0}-\Lambda r^{2}+2\eta q^{2}\ln \left( \frac{r}{l}\right) ,
\label{g(r)}
\end{equation}%
where $m_{0}$ is integration constant related to the mass of the black hole.

We calculate the Ricci and Kretschmann scalars for the solution (\ref{g(r)})
and get them as 
\begin{eqnarray}
R &=&6\Lambda -\frac{2\eta q^{2}}{r^{2}}, \\
&&  \notag \\
R_{\alpha \beta \gamma \delta }R^{\alpha \beta \gamma \delta } &=&12\Lambda
^{2}-\frac{8\Lambda \eta q^{2}}{r^{2}}+\frac{12\eta ^{2}q^{4}}{r^{4}},
\end{eqnarray}%
which show that there is a curvature singularity located at $r=0$, i.e., 
\begin{eqnarray}
\lim_{r\longrightarrow 0}R &\longrightarrow &\infty ,  \notag \\
&&  \notag \\
\lim_{r\longrightarrow 0}R_{\alpha \beta \gamma \delta }R^{\alpha \beta
\gamma \delta } &\longrightarrow &\infty ,
\end{eqnarray}%
and also it is finite for $r\neq 0$. The asymptotical behavior of them are
given by 
\begin{eqnarray}
\lim_{r\longrightarrow \infty }R &\longrightarrow &6\Lambda ,  \notag \\
&&  \notag \\
\lim_{r\longrightarrow \infty }R_{\alpha \beta \gamma \delta }R^{\alpha
\beta \gamma \delta } &\longrightarrow &12\Lambda ^{2},
\end{eqnarray}%
which indicates the spacetime is independent of $\eta $, and it will be
asymptotically (A)dS.

\begin{figure}[tbph]
\centering
\includegraphics[width=0.35\linewidth]{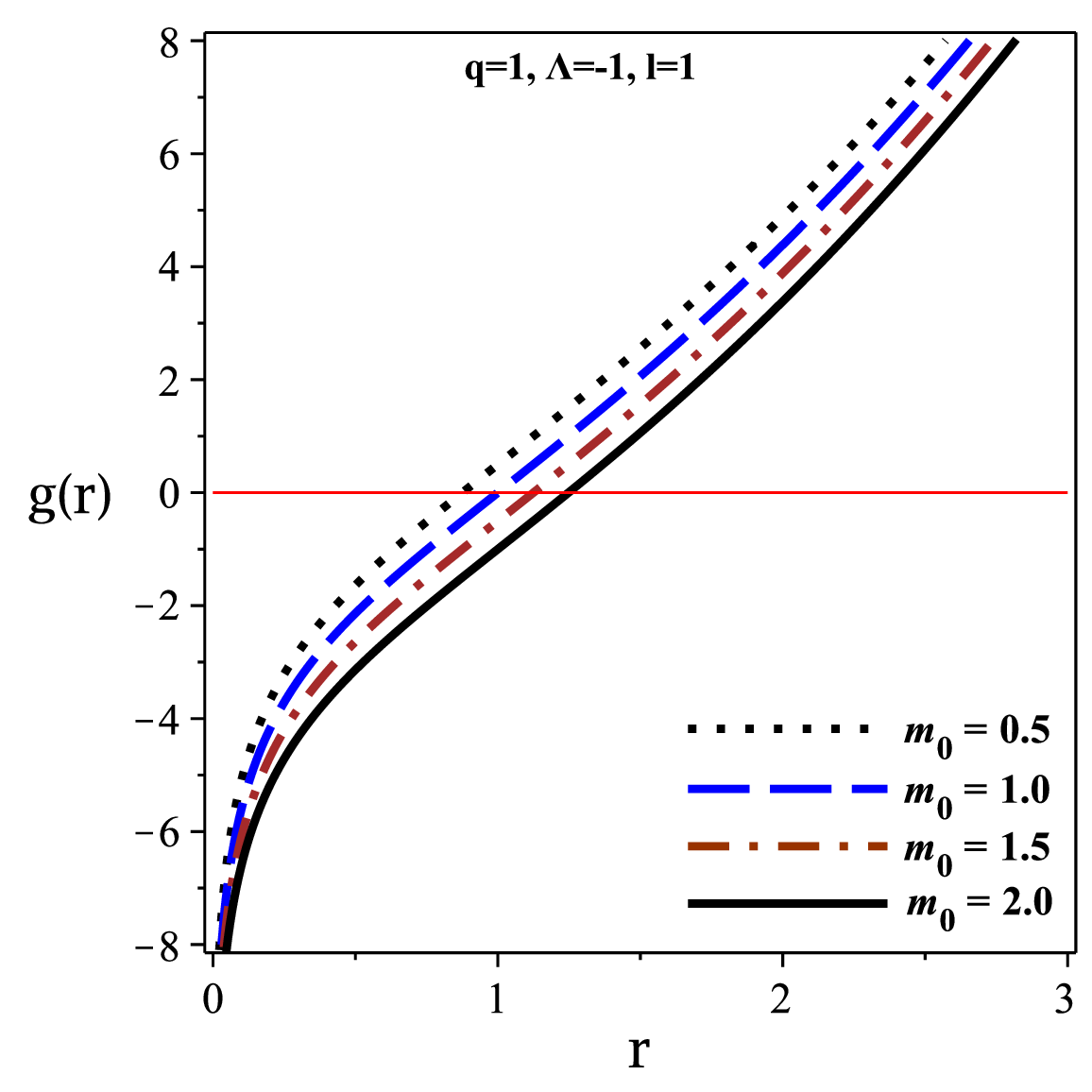} \includegraphics[width=0.35%
\linewidth]{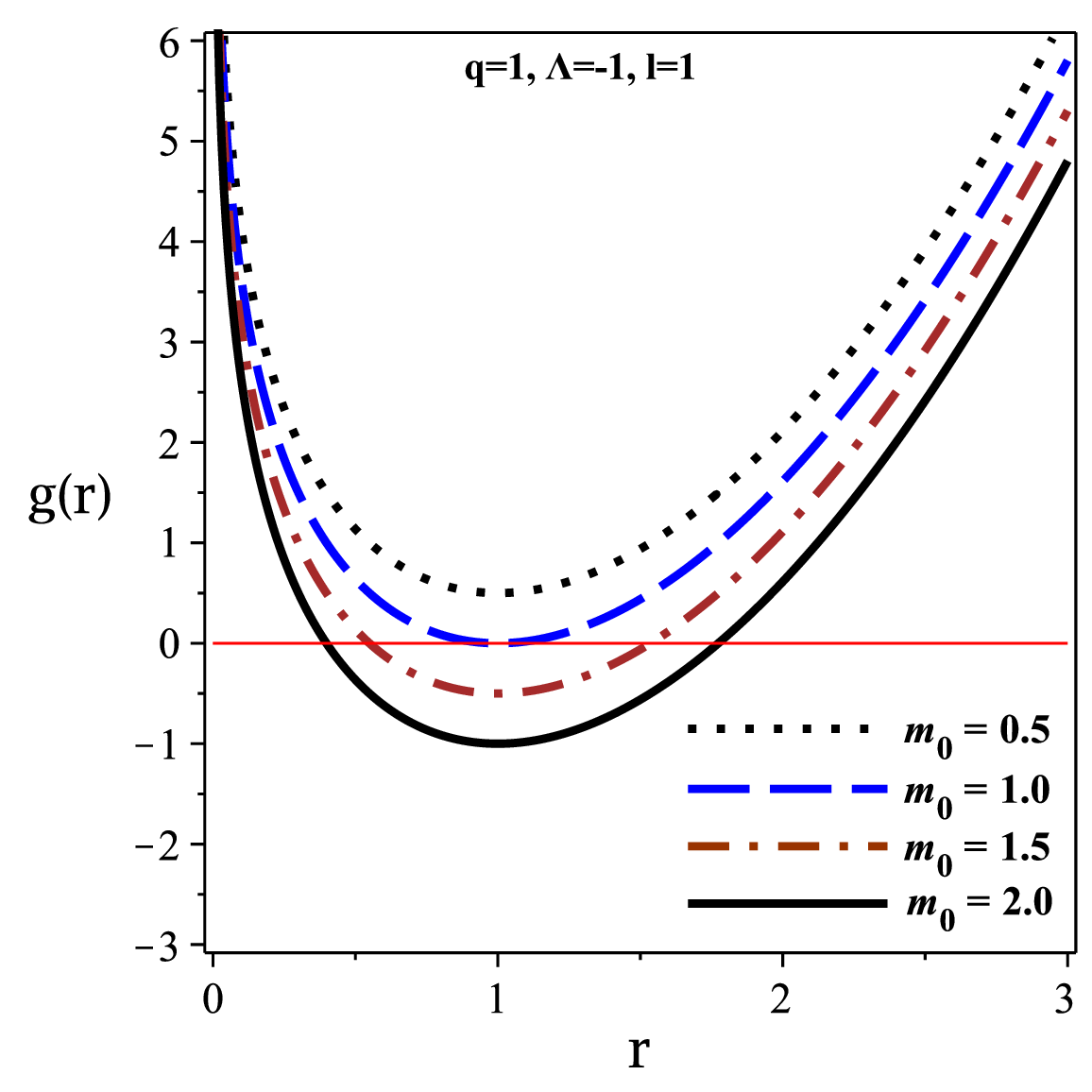}\newline
\caption{The metric function $g(r)$ versus $r$ for different values of the
mass. Left panels for phantom ($\protect\eta =1$), and right panels for
Maxwell case ($\protect\eta =-1$).}
\label{Fig1}
\end{figure}

\begin{figure}[tbph]
\centering
\includegraphics[width=0.35\linewidth]{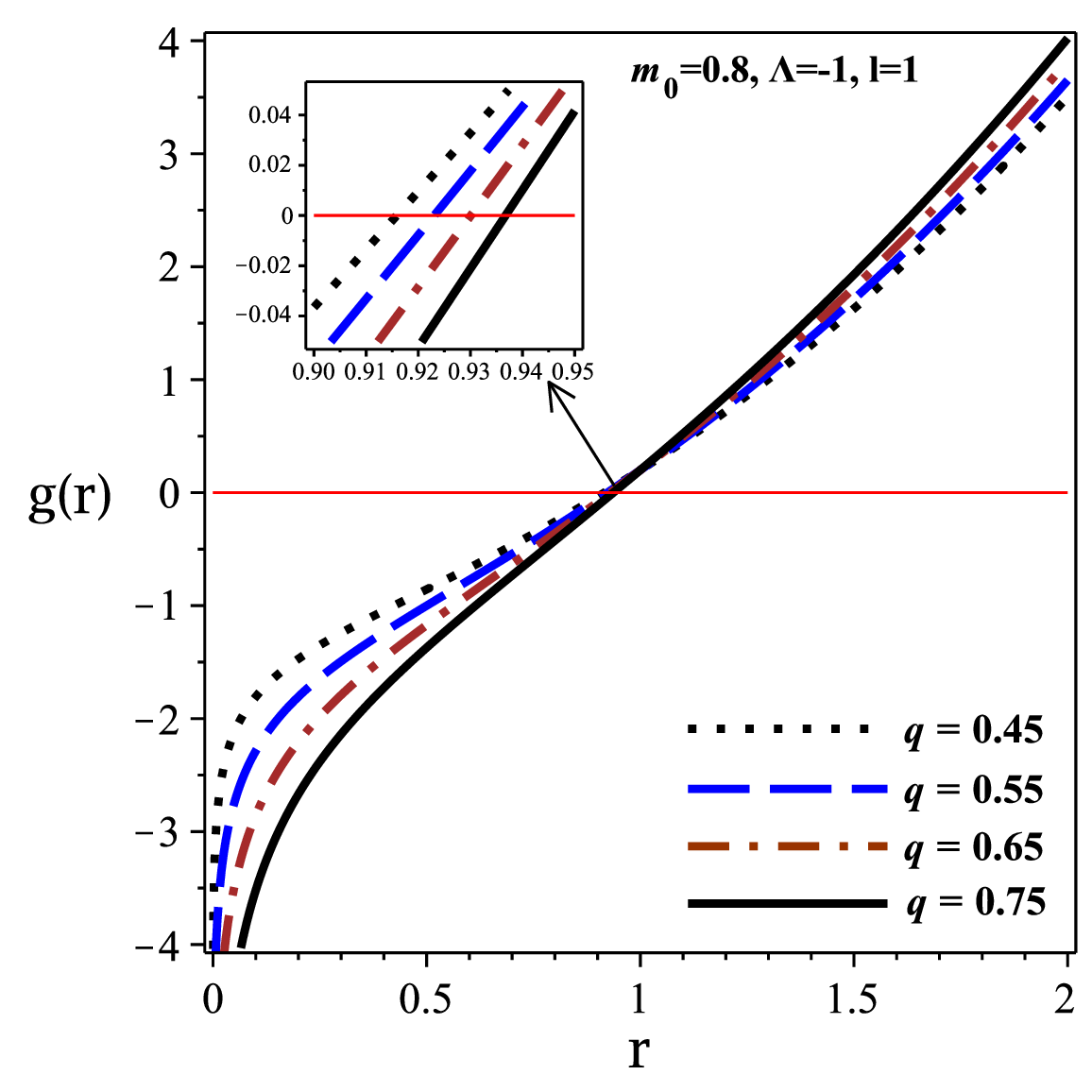} \includegraphics[width=0.35%
\linewidth]{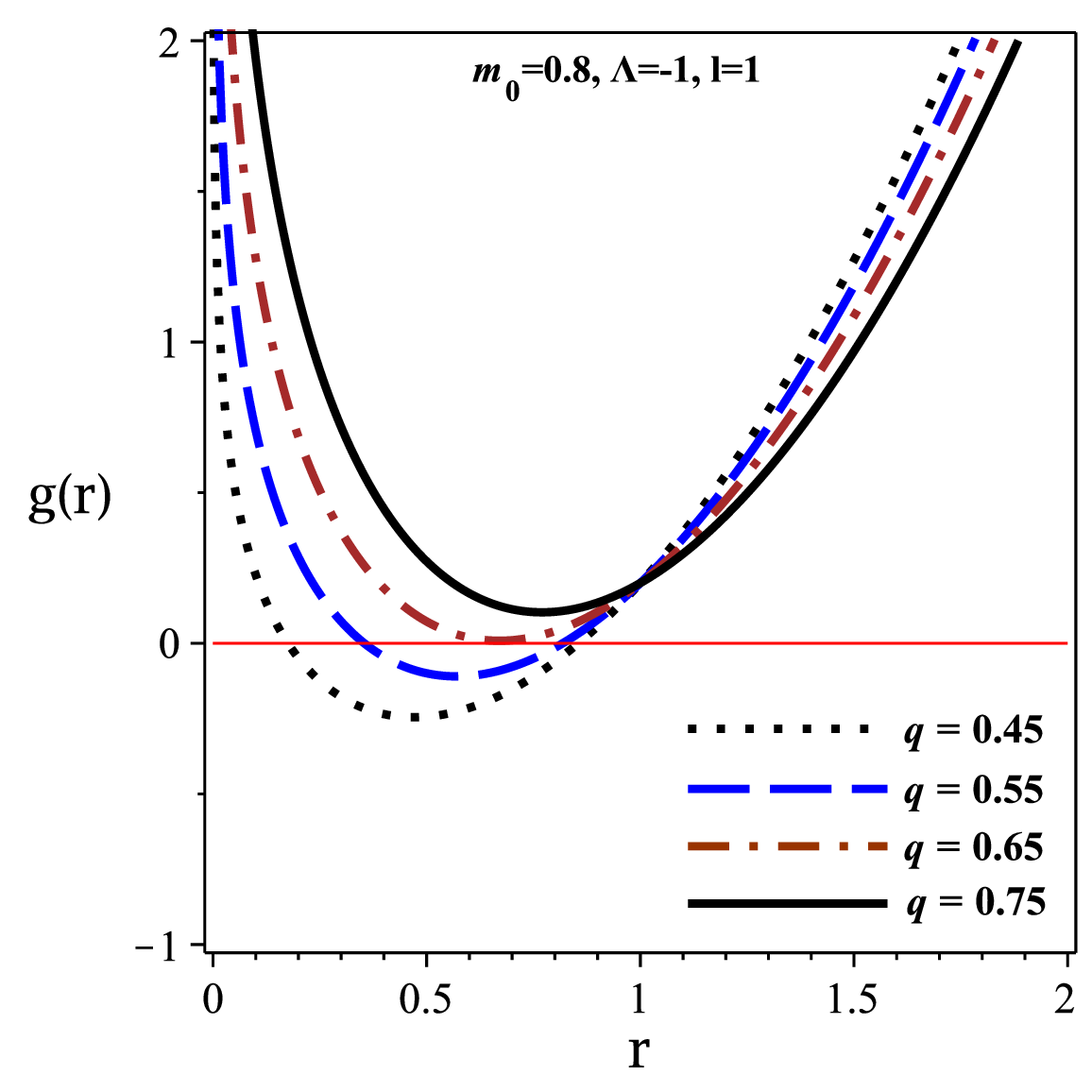}\newline
\caption{The metric function $g(r)$ versus $r$ for different values of the
electrical charge. Left panels for phantom ($\protect\eta =1$), and right
panels for Maxwell case ($\protect\eta =-1$).}
\label{Fig2}
\end{figure}

\begin{figure}[tbph]
\centering
\includegraphics[width=0.35\linewidth]{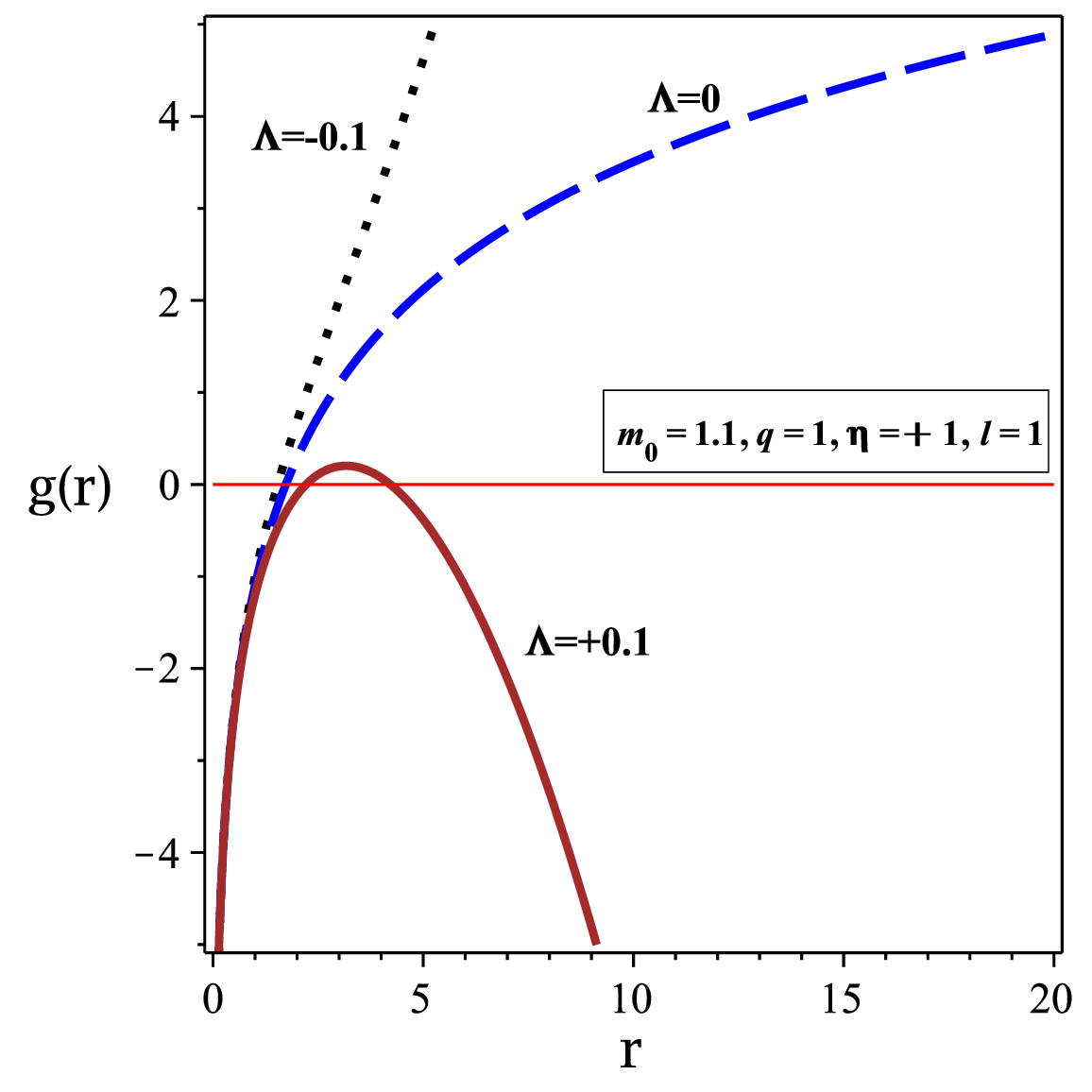} \includegraphics[width=0.35%
\linewidth]{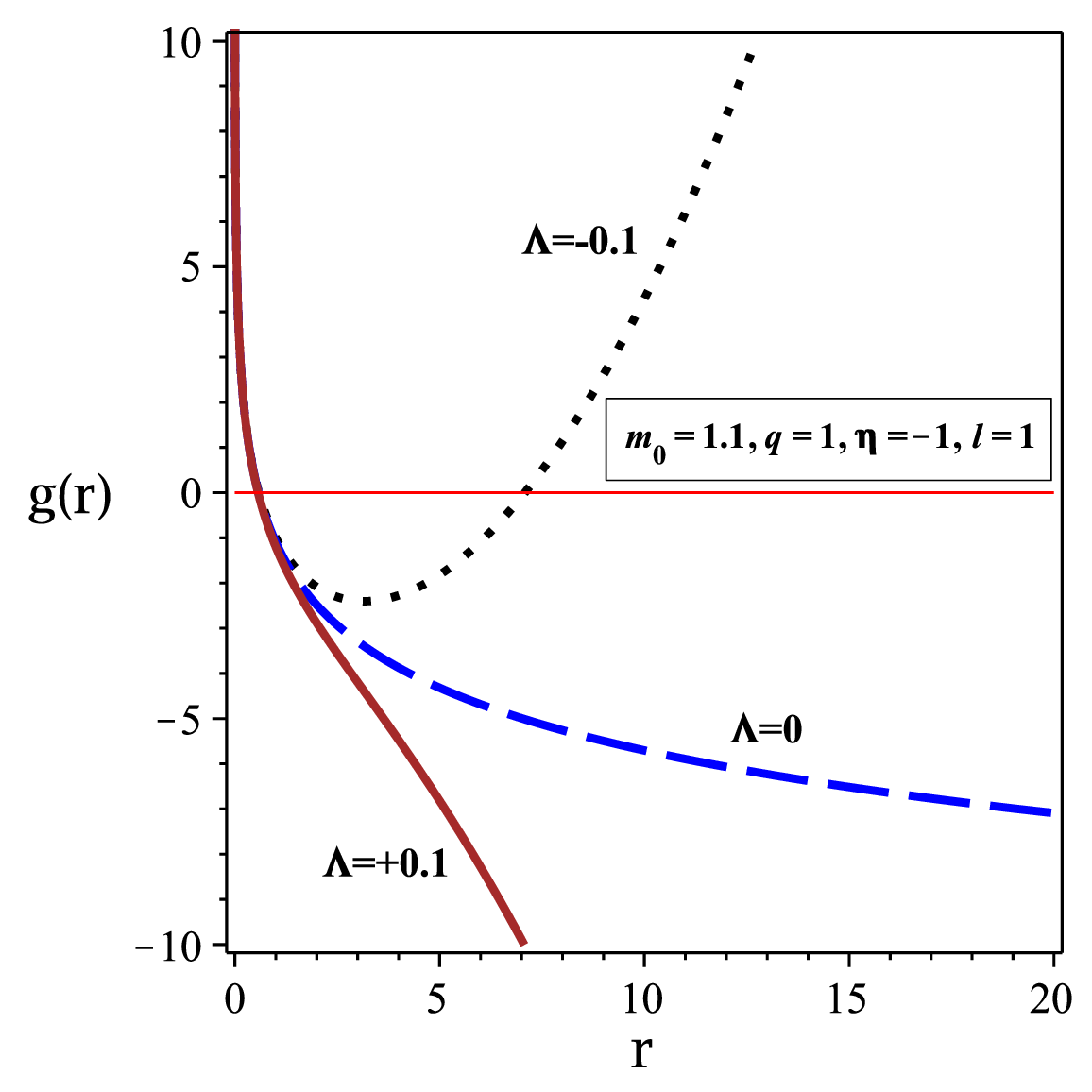}\newline
\caption{The metric function $g(r)$ versus $r$ for different values of the
cosmological constant. Left panels for phantom ($\protect\eta =1$), and
right panels for Maxwell case ($\protect\eta =-1$).}
\label{Fig3}
\end{figure}

\textbf{Horizon Structure:} To evaluate the effects of different parameters
of these black holes, we plot the metric function (\ref{g(r)}) versus $r$ in
Fig. \ref{Fig1}-\ref{Fig3}. This figure gives us some information about the
behavior of horizons. In addition, by studying Figs. \ref{Fig1}-\ref{Fig3},
we can compare the BTZ black holes in the presence of Maxwell and phantom
(anti-Maxwell) fields together.

Considering some values of parameters of this solution, we want to
investigate the effects of mass, the electrical charge, and the cosmological
constant on BTZ black holes in the presence of Maxwell and phantom fields.

\textbf{Mass:} Here, we investigate the impact of the mass parameter on the
event horizon. The findings reveal that as the mass increases, both types of
black holes become larger. Indeed, the radii of black holes increase as $%
m_{0}$ increases (see Fig. \ref{Fig1}). However, there is a significant
distinction between charged BTZ black holes in the phantom case in terms of
the number of roots. In other words, the phantom BTZ black holes exhibit
only one root (or event horizon), as shown in the left panel of Fig. \ref%
{Fig1}. Conversely, BTZ black holes in the presence of the Maxwell field
have two roots (inner and outer roots), one root (extremal black holes), and without root (naked singularities), as illustrated in the right panel of
Fig. \ref{Fig1}. This discrepancy arises from the presence of the phantom
effect.

\textbf{Electrical Charge:} Another interesting result is related to the
effect of the electrical charge on the event horizon. Our findings show that
the higher-charged phantom BTZ black holes have large radii. In other words,
by increasing the electrical charge ($q$), the radius of the phantom black
hole increases (see the left panel in Fig. \ref{Fig2}). But in the presence of Maxwell field when the electrical charge increases, we first encounter
with small BTZ black holes and then the number of roots decreases (see the right panel in Fig. \ref{Fig2}). 

\textbf{Cosmological Constant:} Phantom BTZ black holes exist for three
cases of the cosmological constant ($\Lambda >0$, $\Lambda <0$, and $\Lambda
=0$). Indeed, there is an event horizon for BTZ black holes in the presence
of a phantom field when the values of the cosmological constant are zero,
negative and positive (see the left panel in Fig. \ref{Fig3}). But, for $%
\Lambda >0$, there are two roots which the smaller root is relsted to the
event horizon and another one is the coslomogical horizon. In other words, the
small black holes can be exist for $\Lambda >0$. However, the charged BTZ
black holes (Maxwell case) can exist when the value of the cosmological constant is only negative (see the right panel in Fig. \ref{Fig3}).

In general, there are three essentially different behaviors of BTZ black holes between Maxwell and phantom fields. i) There is only one root for phantom BTZ black holes. However, the roots of BTZ black holes change in the presence of the Maxwell field. ii) The radius of the phantom BTZ black hole increases with the increase of electric charge, which is different from the Maxwell case. Moreover, the number of roots decreases as the electric charge increases. In fact, BTZ black holes can have two roots (inner and outer roots), one root (extremal black holes) and no root (naked singularities) as the electric charge increases. iii) Phantom BTZ black holes can exist for $\lambda <0$, $\Lambda >0$, and $\lambda =0$. However, the BTZ black hole with Maxwell field only exists in the case $\Lambda <0$.

\section{Thermodynamics}

Now, we are interested in to calculate the conserved and thermodynamic
quantities of BTZ black holes in the presence of Maxwell and phantom fields
to check the first law of thermodynamics.

Using the Hawking temperature, which is given by $T=\frac{\kappa }{2\pi }$
(where $\kappa $ is the superficial gravity), we can obtain the Hawking
temperature of phantom BTZ black holes. For this purpose, by considering $%
g(r)=0$, we first express the mass ($m_{0}$) in terms of the radius of the
event horizon ($r_{+}$), the cosmological constant and the charge $q$ in the
following form 
\begin{equation}
m_{0}=2\eta q^{2}\ln \left( \frac{r_{+}}{l}\right) -\Lambda r_{+}^{2},
\label{m0}
\end{equation}

Then, we calculate the superficial gravity for the mentioned spacetime (\ref%
{Metric}), which leads to 
\begin{equation}
\kappa =\left. \frac{g^{\prime }(r)}{2}\right\vert _{r=r_{+}},  \label{kGR}
\end{equation}%
by considering BTZ black holes (\ref{g(r)}), and by substituting the mass (%
\ref{m0})\ within Eq. (\ref{kGR}), one can calculate the superficial gravity
as 
\begin{equation}
\kappa =\frac{\eta q^{2}}{r_{+}}-\Lambda r_{+},
\end{equation}%
and so the Hawking temperature leads to 
\begin{equation}
T=\frac{\eta q^{2}}{2\pi r_{+}}-\frac{\Lambda r_{+}}{2\pi }.  \label{TemGR}
\end{equation}

The electric charge of a black hole can be obtained by using the Gauss law
in the following form 
\begin{equation}
Q=\frac{1}{4\pi }\left. \int_{0}^{2\pi }F_{tr}\left( r\right) \sqrt{g}%
d\varphi \right\vert _{r=r_{+}}=\frac{q}{2}  \label{QGR}
\end{equation}%
where $F_{tr}\left( r\right) =\frac{q}{r}$, and $g=det(g_{\mu \nu })=r^{2}$.

Using $F_{\mu \nu }=\partial _{\mu }A_{\nu }-\partial _{\nu }A_{\mu }$, one
can find the nonzero component of the gauge potential in which is $%
A_{t}=-\int F_{tr}\left( r\right) dr$. So, we can get the electric potential
($U$) at the event's horizon with respect to the reference ($%
r\rightarrow\infty $) as 
\begin{equation}
U=-\int_{r_{+}}^{+\infty }F_{tr}\left( r\right) dr=q\ln \left( \frac{r_{+}}{l%
}\right) .  \label{elcpoGR}
\end{equation}

To obtain the entropy of BTZ black holes, one can use of the area law 
\begin{equation}
S=\frac{A}{4},  \label{SGR}
\end{equation}%
where $A$ is the horizon area and is defined by 
\begin{equation}
A=\left. \int_{0}^{2\pi }\sqrt{g_{\varphi \varphi }}d\varphi \right\vert
_{r=r_{+}}=\left. 2\pi r\right\vert _{r=r_{+}}=2\pi r_{+},  \label{AGR}
\end{equation}%
where $g_{\varphi \varphi }=r^{2}$. So, the entropy of BTZ black holes in
the presence Maxwell and phantom fields is given by

\begin{equation}
S=\frac{\pi r_{+}}{2}.  \label{EntGR}
\end{equation}

Another interesting quantity is related to the total mass of the black hole.
To obtain the total mass we use of the Ashtekar-Magnon-Das (AMD) approach 
\cite{AMDI,AMDII}, and we have 
\begin{equation}
M=\frac{m_{0}}{8},  \label{AMDMassGR}
\end{equation}%
substituting the mass (\ref{m0}) within the equation (\ref{AMDMassGR}),
yields 
\begin{equation}
M=\frac{\eta q^{2}}{4}\ln \left( \frac{r_{+}}{l}\right) -\frac{\Lambda
r_{+}^{2}}{8}.  \label{Mr}
\end{equation}

\begin{figure}[tbph]
\centering
\includegraphics[width=0.35\linewidth]{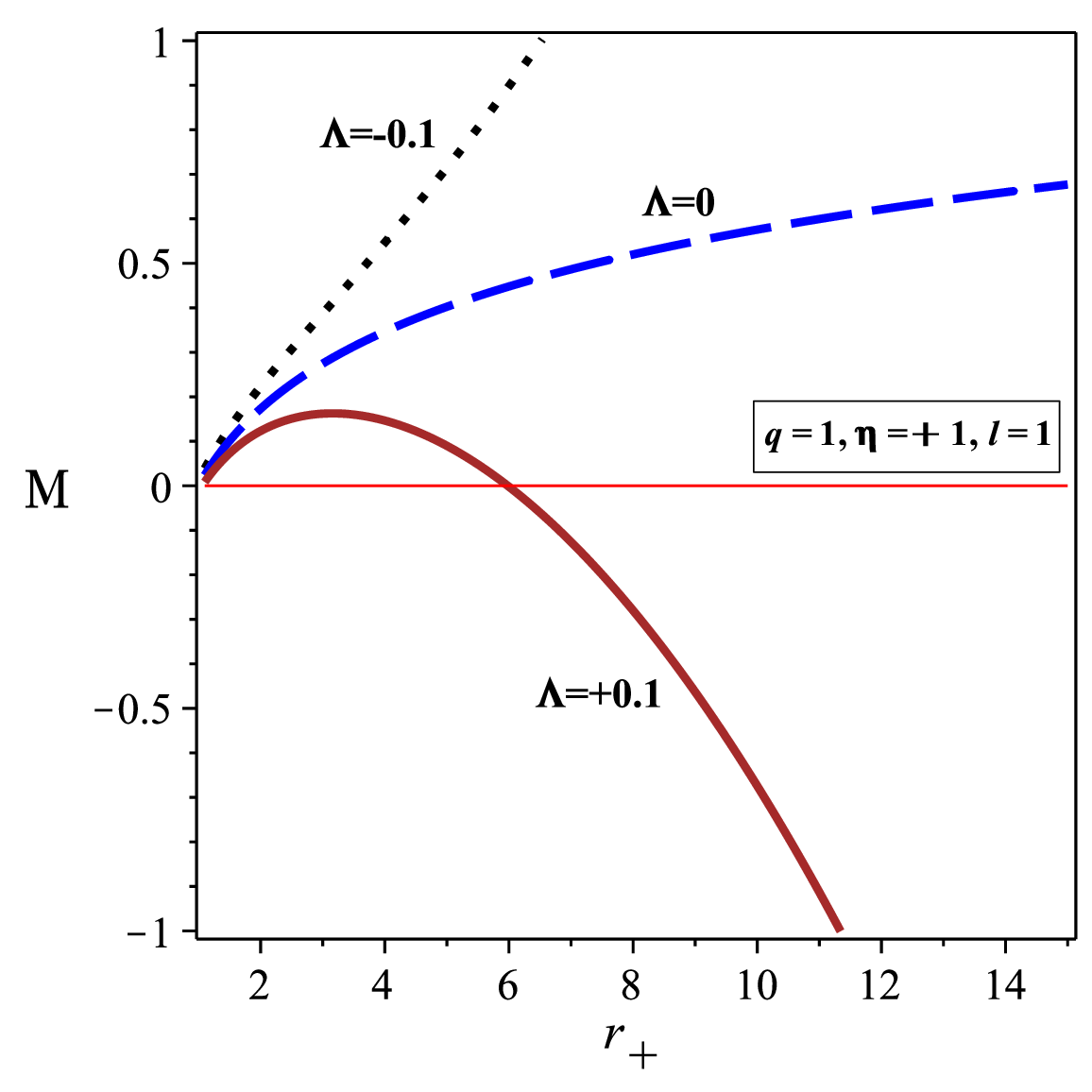} \includegraphics[width=0.35%
\linewidth]{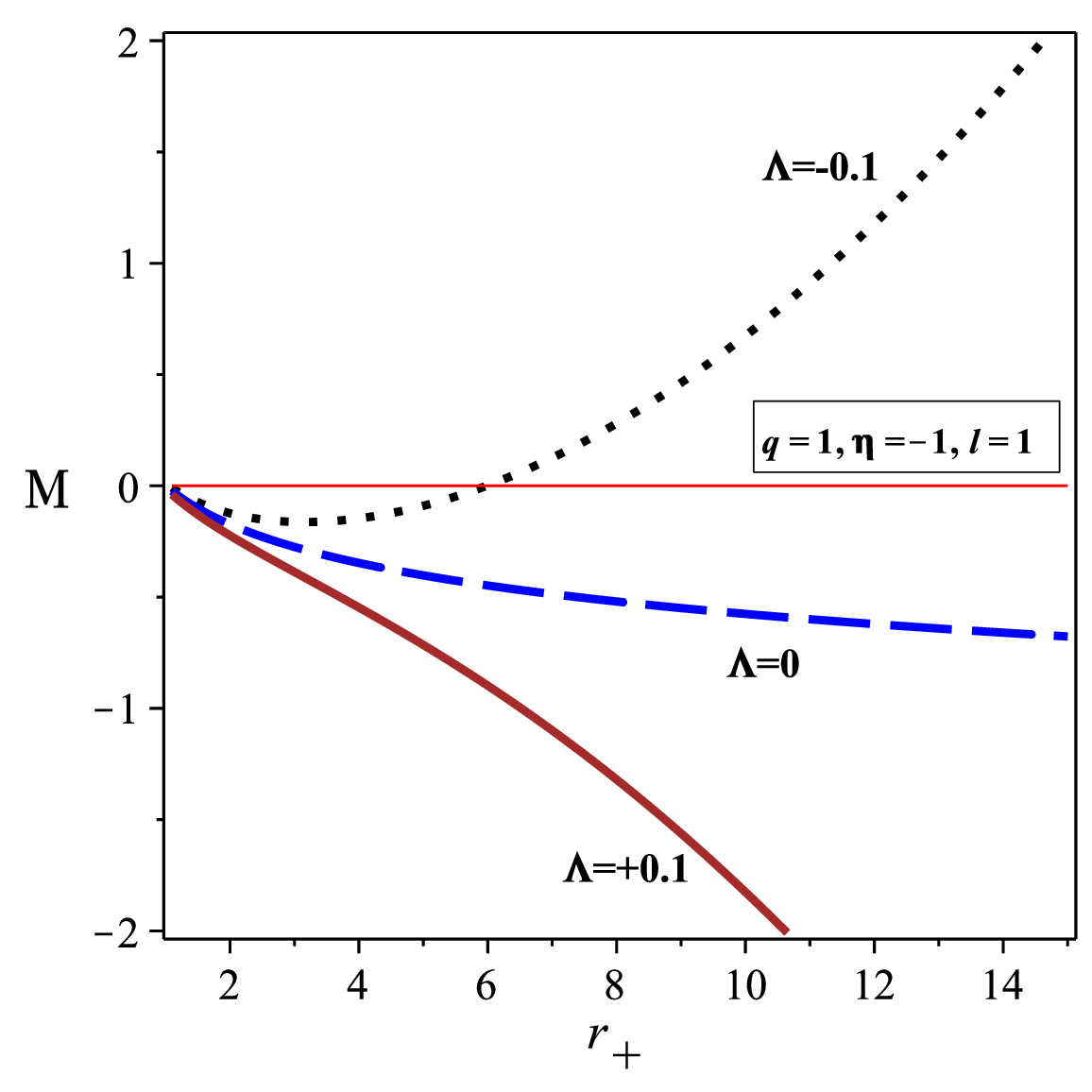}\newline
\caption{$M$ versus $r_{+}$ for different values of the cosmological
constant. Left panels for phantom ($\protect\eta =1$), and right panels for
Maxwell case ($\protect\eta =-1$).}
\label{Fig4}
\end{figure}

To have a positive mass for the phantom case ($\eta > 0$), the cosmological constant can be positive, negative, or zero, as long as $\Lambda < \frac{2\eta q^{2}}{r_{+}^{2}}\ln \left( \frac{r_{+}}{l}\right)$ holds. This
constraint indicates that the mass of phantom BTZ black holes is always
positive when $\Lambda < 0$ and $\Lambda = 0$ (see dotted and dashed lines
in the left panel of Fig. \ref{Fig4}). However, for $\Lambda > 0$ to have a
positive mass, we need to satisfy $\Lambda < \frac{2\eta q^{2}}{r_{+}^{2}}%
\ln \left( \frac{r_{+}}{l}\right)$. This reveals that the total mass of
black holes cannot be positive for large phantom dS BTZ black holes (see the
continuous line in the left panel of Fig. \ref{Fig4}). On the other hand,
the mass of the Maxwell AdS BTZ black hole with a large radius is positive
only for a negative cosmological constant (see the dotted line in the right
panel of Fig. \ref{Fig4}). This result is consistent with the behavior of
the mass ($m_{0}$) for the metric function obtained (compare Fig. \ref{Fig3}
with Fig. \ref{Fig4}).

It is straightforward to show that the conserved and thermodynamics
quantities satisfy the first law of thermodynamics in the following form 
\begin{equation}
dM=TdS+\eta UdQ,
\end{equation}%
and one can define the intensive parameters conjugate to $S$ and $Q$. These
quantities are the electric potential $\left( U=\left( \frac{\partial M}{%
\partial Q}\right) _{S}\right) $, and the temperature $\left( T=\left( \frac{%
\partial M}{\partial S}\right) _{Q}\right) $ which are the same as those
calculated for the electric potential (\ref{elcpoGR}), and the temperature (%
\ref{TemGR}).

\section{\textbf{Thermal Stability}}

\subsection{Local Stability}

Heat capacity can study a thermodynamic system's local stability in the
canonical ensemble context. Therefore, we evaluate the heat capacity to find
the local stability for phantom BTZ black holes. Indeed, we study the effect
of phantom field on local stability of three-dimensional black holes and
compare it with Maxwell field. The heat capacity is given by 
\begin{equation}
C_{Q}=\frac{T}{\left( \frac{\partial T}{\partial S}\right) _{Q}}.
\end{equation}

By replacing Eq. (\ref{EntGR}) in Eq. (\ref{TemGR}), we can find the Hawking
temperature ($T$) versus $S$ in the following form 
\begin{equation}
T=\frac{\eta Q}{S}-\frac{\Lambda S}{\pi ^{2}},  \label{TS}
\end{equation}%
and after some calculation, we can extract the heat capacity of phantom BTZ
black holes, which leads to 
\begin{equation}
C_{Q}=\frac{\left( \Lambda S^{2}-\eta \pi ^{2}Q^{2}\right) S}{\Lambda
S^{2}+\eta \pi ^{2}Q^{2}},  \label{C}
\end{equation}

We solve $C_{Q}$ to find the roots of the heat capacity, which are 
\begin{equation}
S_{root_{1}}=\frac{\pi Q\sqrt{\eta \Lambda }}{\Lambda },~~~\&~~~S_{root_{2}}=%
\frac{-\pi Q\sqrt{\eta \Lambda }}{\Lambda }.
\end{equation}

\begin{figure}[tbph]
\centering
\includegraphics[width=0.35\linewidth]{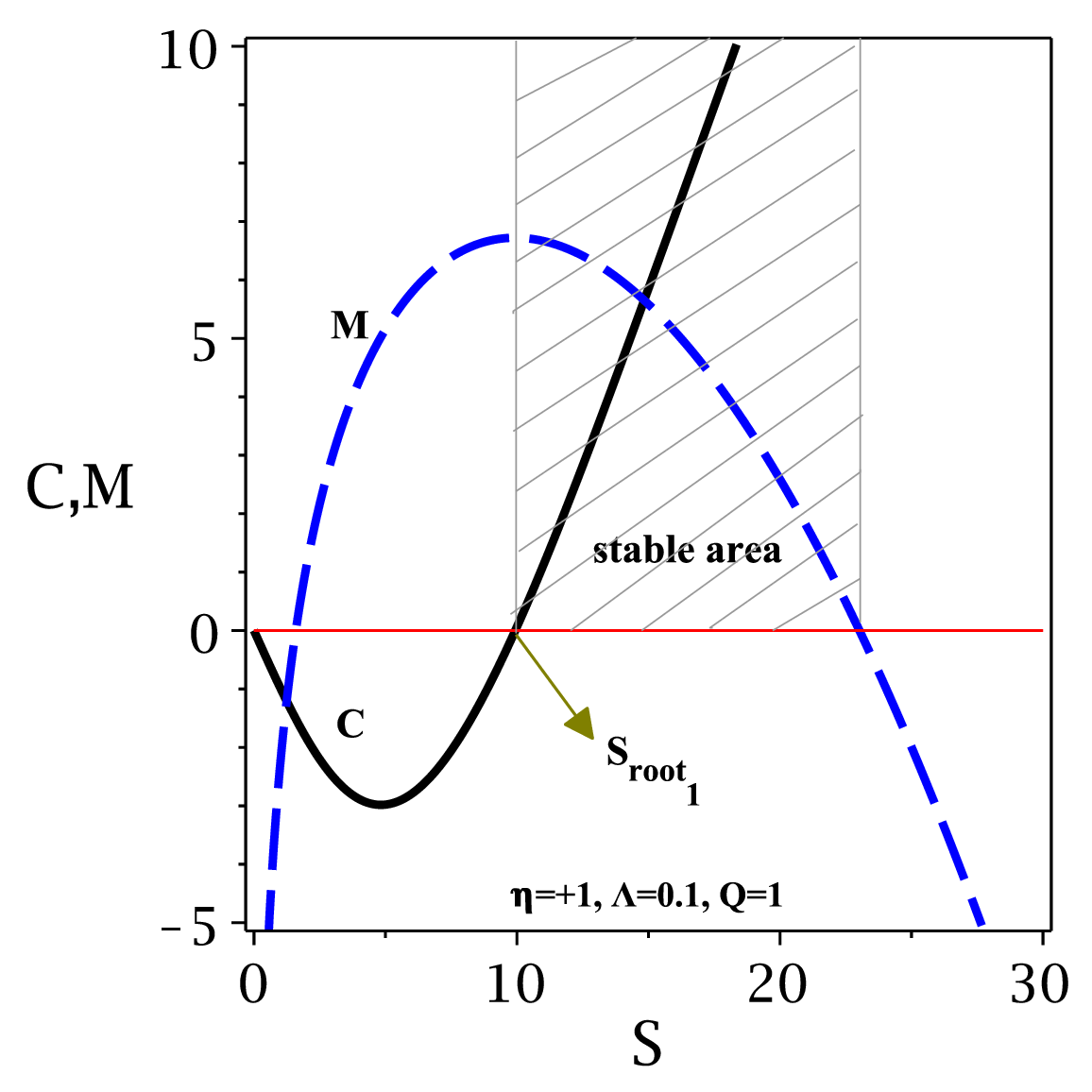} \includegraphics[width=0.35%
\linewidth]{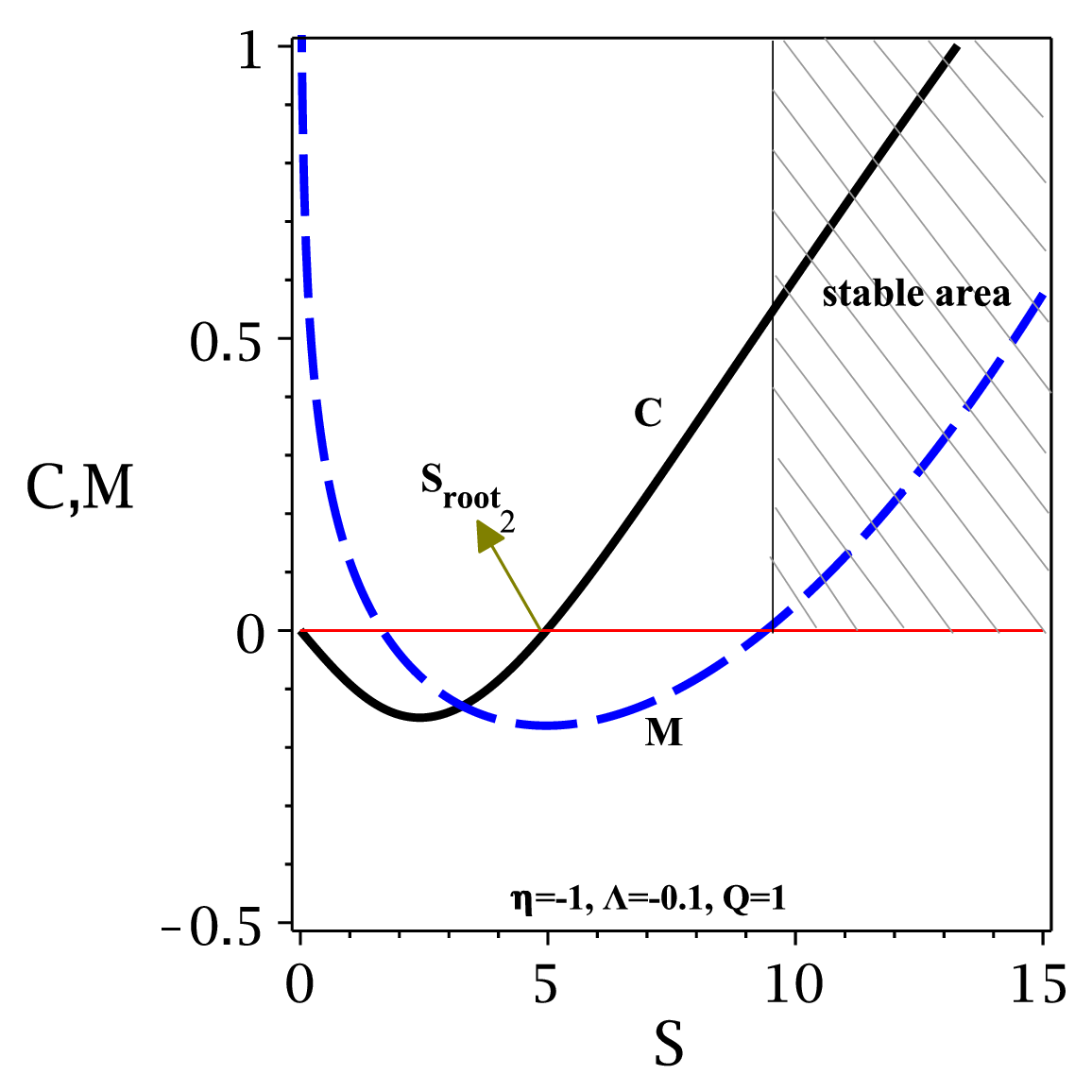}\newline
\caption{The heat capacity $C_{Q}$ and mass ($M$) versus $S$ for different
values of the mass. Left panel for phantom ($\protect\eta =1$), and right
panel for Maxwell case ($\protect\eta =-1$).}
\label{Fig5}
\end{figure}

To have real root, we must respect to the condition $\eta \Lambda >0$. This
imposes the following conditions for phantom and Maxwell fields, which are:

\textbf{Phantom case (}$\eta =1$\textbf{):} to have a positive value of $%
\eta \Lambda $ (i.e., $\eta \Lambda >0$), the cosmological constant must be
positive. In this case, $S_{root_{1}}>0$ and $S_{root_{2}}<0$. Furthermore,
there are two roots for the mass. The mass is positive between these roots.
Our analysis reveals that the phantom dS BTZ black hole satisfies local
stability when it is between $S_{root_{1}}$ of the heat capacity and the
second root of the mass. In other words, medium phantom black holes with a
positive value of the cosmological constant (i.e. phantom dS black holes)
justify the local stability condition because the mass and the heat capacity
are positive (see the hatched area in the left panel of Fig. \ref{Fig5}).

\textbf{Maxwell case (}$\eta =-1$\textbf{):} the cosmological constant is
negative ($\Lambda <0$) when we consider the condition $\eta \Lambda >0$ for
the Maxwell case. Therefore, $S_{root_{1}}<0$, and $S_{root_{2}}>0$. In
other words, only $S_{root_{2}}$ represents the real root of the heat
capacity. Our findings indicate that charged AdS BTZ black holes with a
large radius (entropy) can be stable. In other words, after the second root
of the mass, both the heat capacity and mass are simultaneously positive.
Therefore, large BTZ black holes with a negative value of the cosmological
constant satisfy local stability (see the hatched area in the right panel of
Fig. \ref{Fig5}).

To find the heat capacity divergence points, we solve the denominator of Eq.
(\ref{C}) in terms of entropy. We extract them in the following forms 
\begin{equation}
S_{{div}_{1}}=\frac{\pi Q\sqrt{-\eta \Lambda }}{\Lambda },~~~\&~~~S_{{div}%
_{2}}=\frac{-\pi Q\sqrt{-\eta \Lambda }}{\Lambda }.
\end{equation}

\begin{figure}[tbph]
\centering
\includegraphics[width=0.35\linewidth]{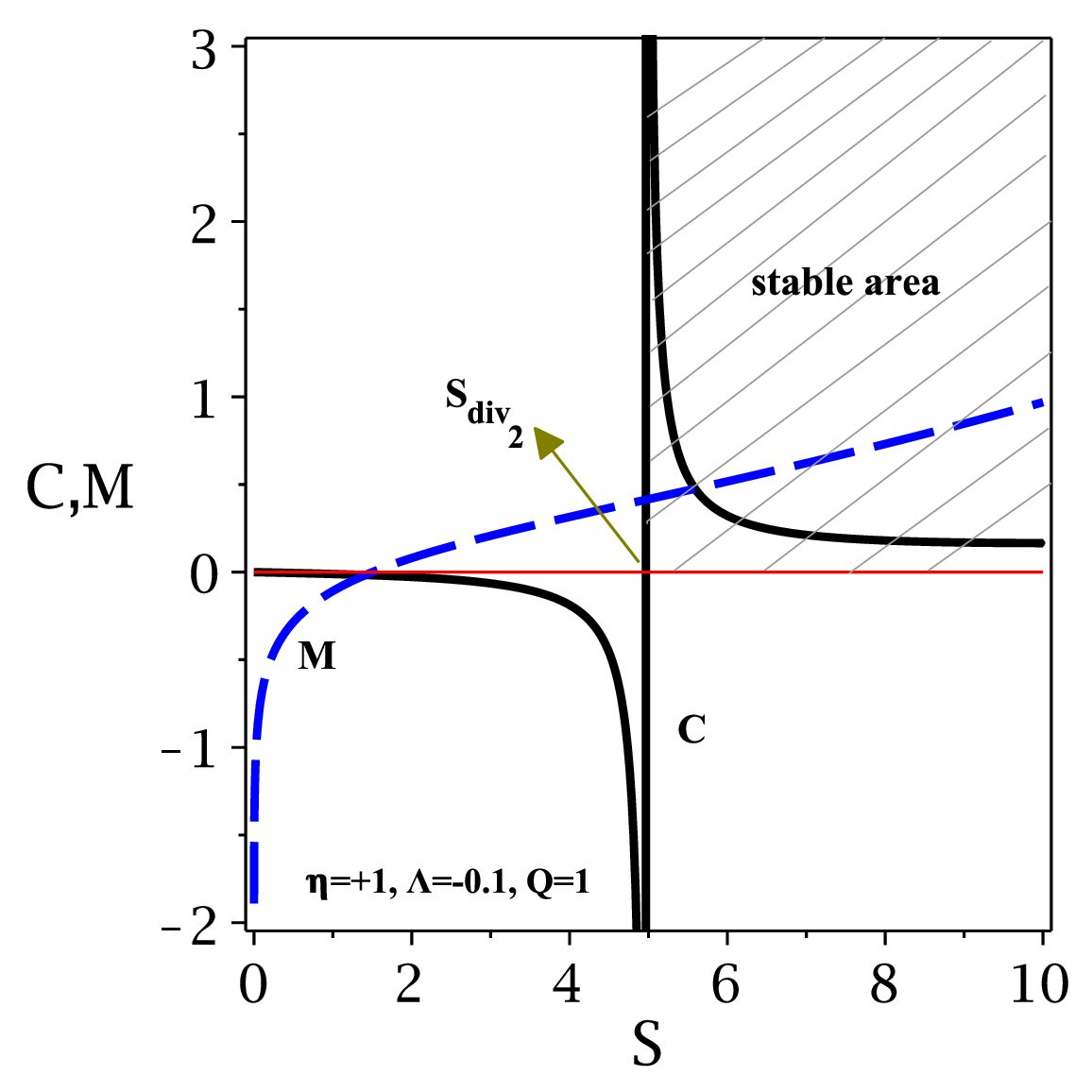} \includegraphics[width=0.35%
\linewidth]{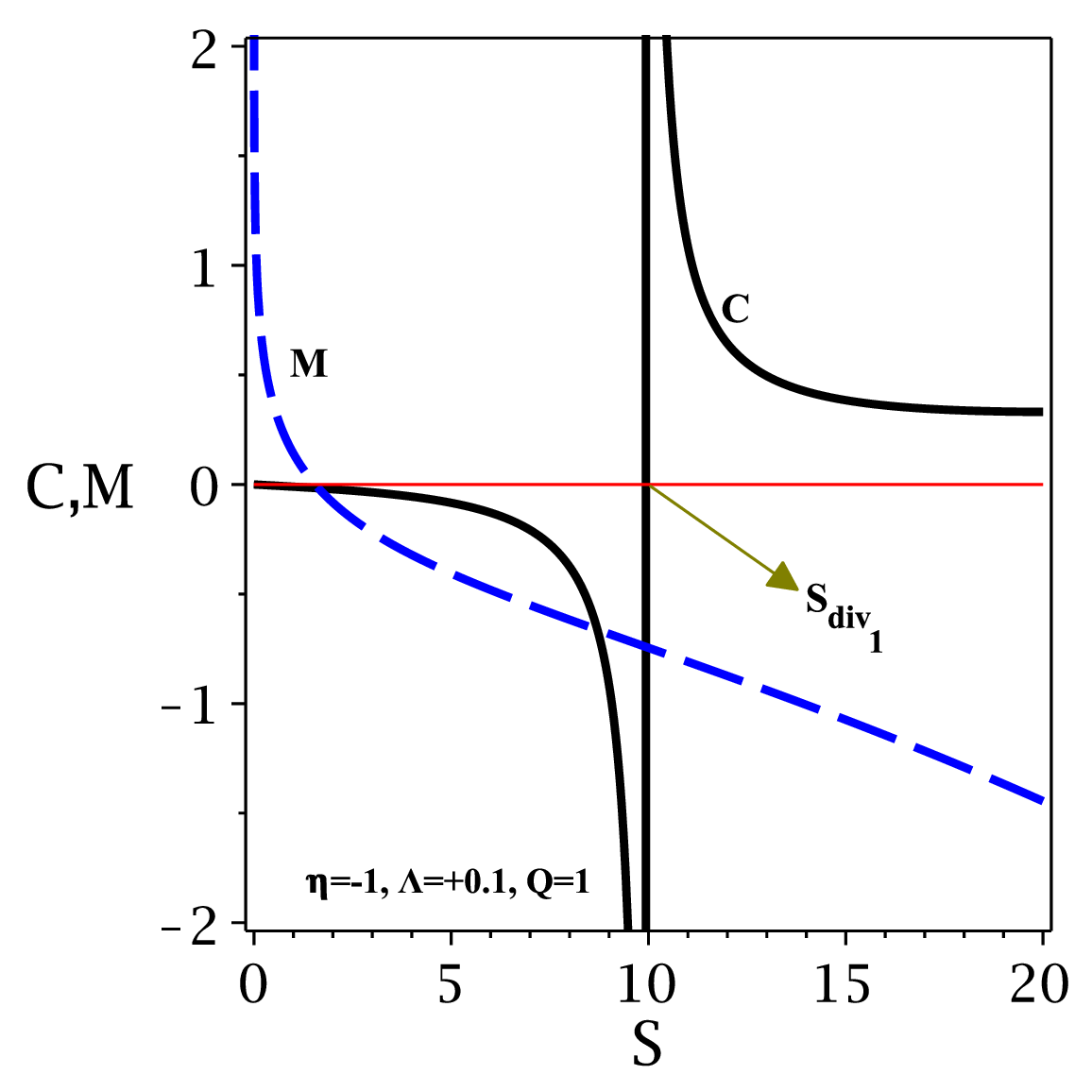}\newline
\caption{The heat capacity $C_{Q}$ and mass ($M$) versus $S$ for different
values of the mass. Left panel for phantom ($\protect\eta =1$), and right
panel for Maxwell case ($\protect\eta =-1$).}
\label{Fig6}
\end{figure}

To have real root, we must respect to the condition $-\eta \Lambda >0$. This
imposes the following conditions for phantom and Maxwell fields, which are:

\textbf{Phantom case (}$\eta =1$\textbf{):} by applying the condition $%
-\eta\Lambda >0$, the cosmological constant must be negative (i.e., $\Lambda
<0$), and it implies that $S_{{div}_{1}}<0$, and $S_{{div}_{2}}>0$.
Therefore, there is a divergence point ($S_{{div}_{2}}$) for the heat
capacity. The behavior of the mass and the heat capacity indicate that
phantom AdS black holes with a large radius (entropy) satisfy local
stability. In other words, large phantom AdS black holes are stable (see the
hatched area in the left panel of Fig. \ref{Fig6}).

\textbf{Maxwell case (}$\eta =-1$\textbf{):} for this case, the cosmological
constant must be positive ($\Lambda >0$) when considering the condition $%
-\eta\Lambda >0$. This condition leads to $S_{{div}_{1}}>0 $ and $S_{{div}%
_{2}}<0$. The results in the right panel of Fig. \ref{Fig6} demonstrate that
there is no local stability region for charged BTZ dS black holes, as the
mass and heat capacity are not simultaneously positive. In other words, the
BTZ black hole in the presence of the Maxwell field with a positive
cosmological constant value cannot satisfy local stability.

Briefly, our results reveal that:

i) for $\Lambda>0$ (dS case), medium phantom BTZ black holes are stable.
However, charged BTZ black holes in the presence of the Maxwell field are
not stable.

ii) for $\Lambda<0$ (AdS case), large black holes in the presence of phantom
and Maxwell fields exhibit local stability. However, there is a distinct
stable region. In other words, phantom BTZ black holes have a larger stable
area compared to the Maxwell case.

\subsection{Global Stability}

Hawking and Page were the first to suggest studying the global stability of
black holes \cite{HawkingP1983}. They proposed that the global stability of
a black hole can be assessed in the grand canonical ensemble by calculating
the Gibbs free energy. A black hole is deemed globally stable if it
possesses negative Gibbs free energy \cite{CaiW2004,DehghaniS2019}. Here,
our objective is to analyze the global stability of phantom BTZ black holes
using the Gibbs free energy approach.

It is notable that in the context of the black holes, the Gibbs free energy
is defined in the following form 
\begin{equation}
G=M(S,Q,\Lambda )-TS-QU.  \label{G0}
\end{equation}

By replacing Eq. (\ref{QGR}), and Eq. (\ref{EntGR}), in Eq. (\ref{Mr}), we
rewrite the mass of the phantom BTZ black holes as 
\begin{equation}
M(S,Q,\Lambda )=\eta Q^{2}\ln \left( \frac{2S}{\pi l}\right) -\frac{\Lambda
S^{2}}{2\pi ^{2}}.  \label{MSQLambda}
\end{equation}

Using Eqs. (\ref{elcpoGR}), (\ref{QGR}), (\ref{TS}), and (\ref{MSQLambda})
in Eq. (\ref{G0}), we can obtain the Gibbs free energy in the following form 
\begin{equation}
G=\left( \eta -2\right) Q^{2}\ln \left( \frac{2S}{\pi l}\right) +\frac{%
\Lambda S^{2}}{2\pi ^{2}}-\eta Q^{2}.  \label{G}
\end{equation}

The global stability areas are deduced from the negative area of the Gibbs free energy (i.e., $G<0$). We apply the constraint $G<0$ to find the global stability areas. This constraint imposes the following two conditions:

\textit{Condition I:} by applying $G<0$ for phantom BTZ black holes ($\eta
=1 $), we find that 
\begin{equation}
\Lambda <\frac{2\pi ^{2}Q^{2}}{S^{2}}\left( \ln \left( \frac{2S}{\pi l}%
\right) +1\right) ,
\end{equation}%
where by adjusting thermodynamic quantities, we can satisfy the above condition. This means that global stability can be achieved for positive, negative, and zero values of the cosmological constant.

\textit{Condition II:} by considering $G<0$ for Maxwell BTZ black holes ($%
\eta =-1$), we can extract the following constraint 
\begin{equation}
\Lambda <\frac{2\pi ^{2}Q^{2}}{S^{2}}\left( 3\ln \left( \frac{2S}{\pi l}%
\right) -1\right) ,
\end{equation}%
so, we can find the global stability areas for $\Lambda >0$, $\Lambda <0$, and $\Lambda =0$ by adjusting the suitable parameters of the above condition.

\begin{figure}[tbph]
\centering
\includegraphics[width=0.35\linewidth]{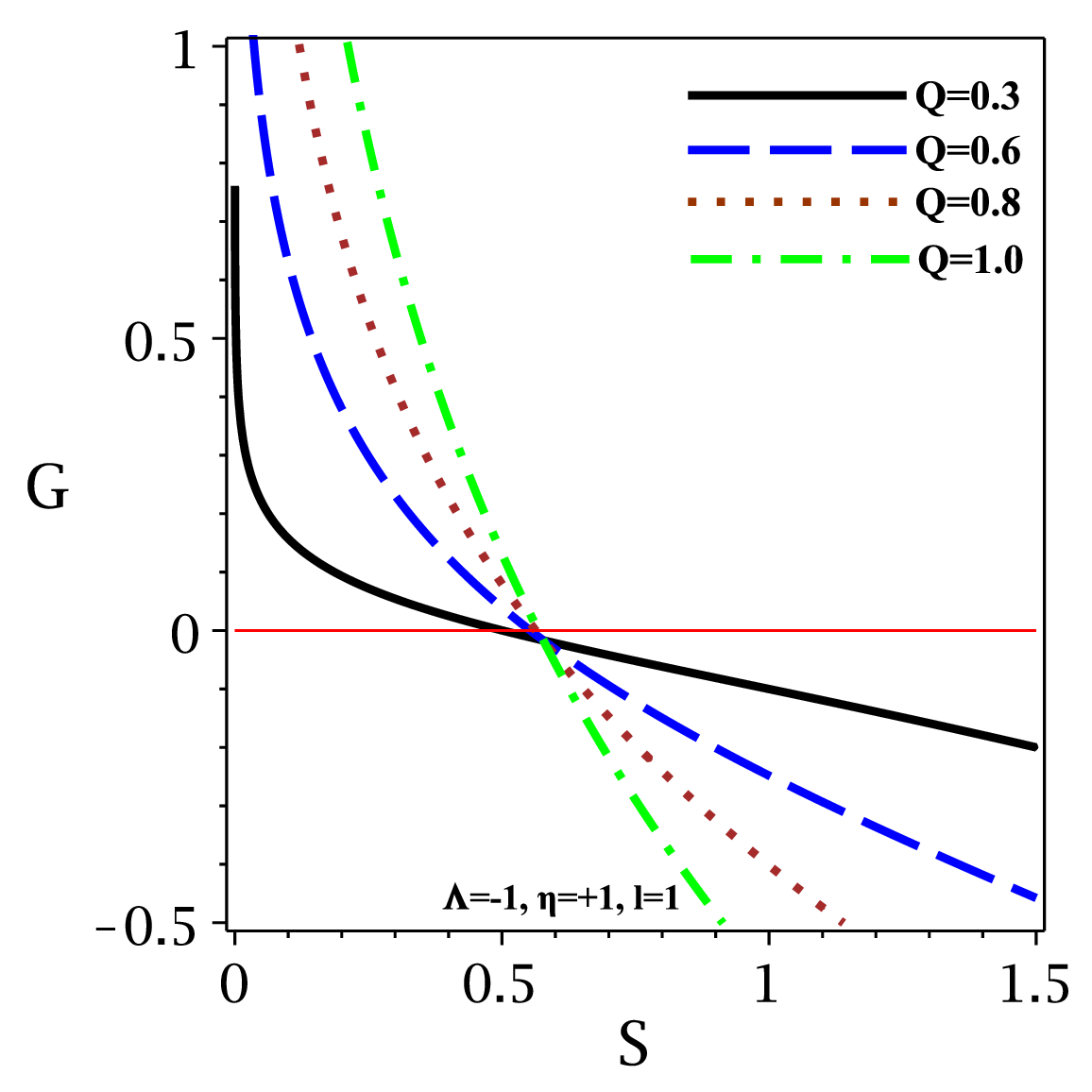} \includegraphics[width=0.35%
\linewidth]{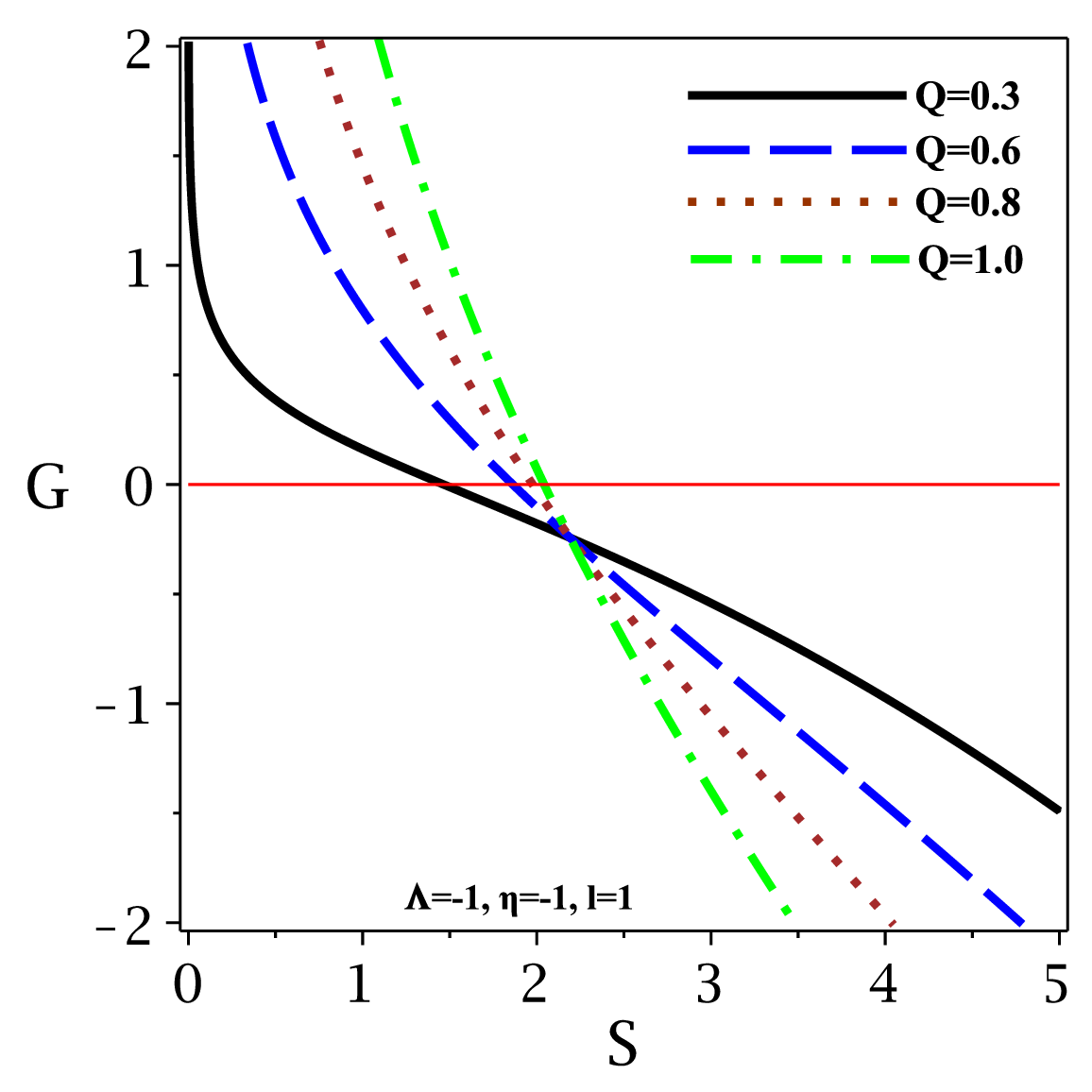} \newline
\includegraphics[width=0.35\linewidth]{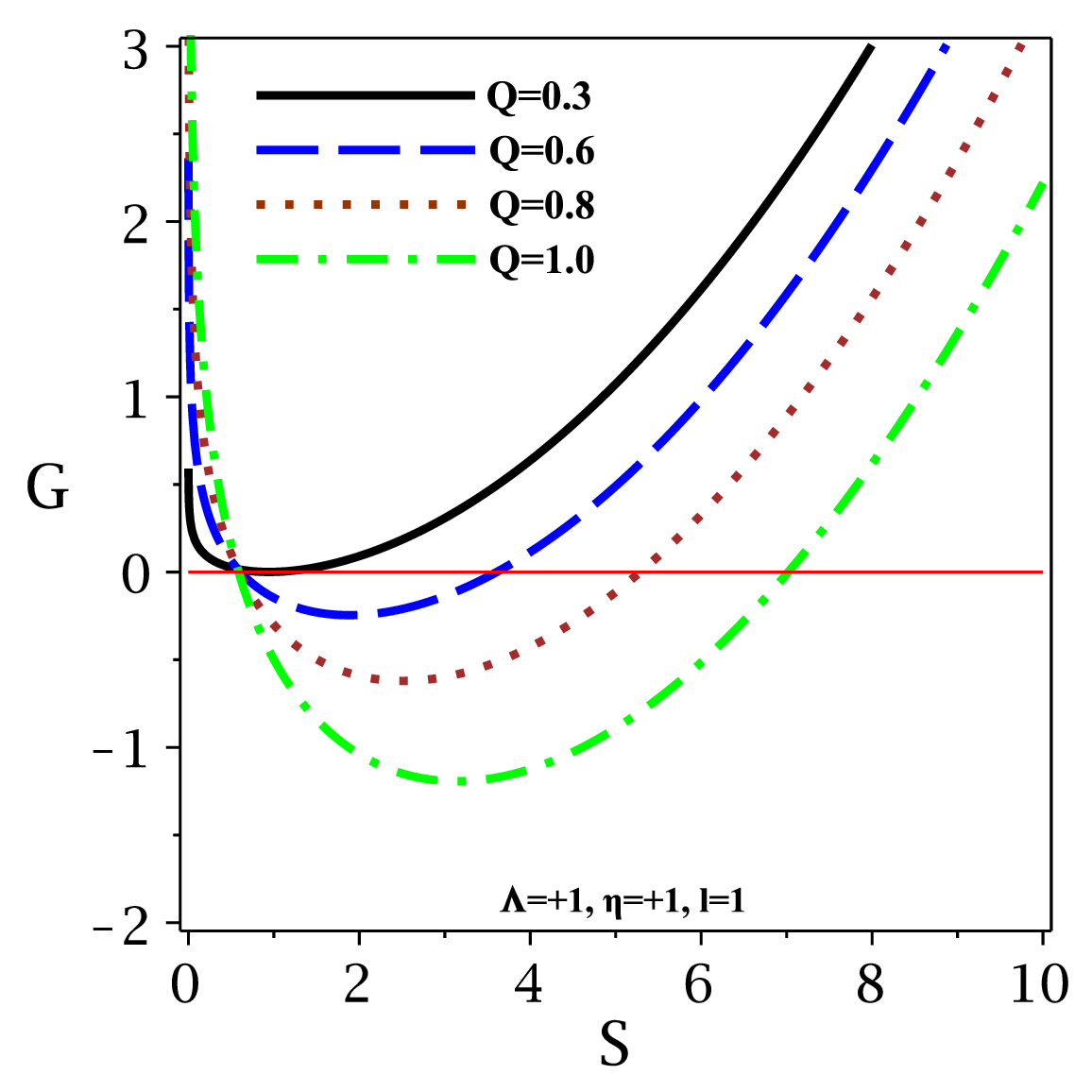} \includegraphics[width=0.35%
\linewidth]{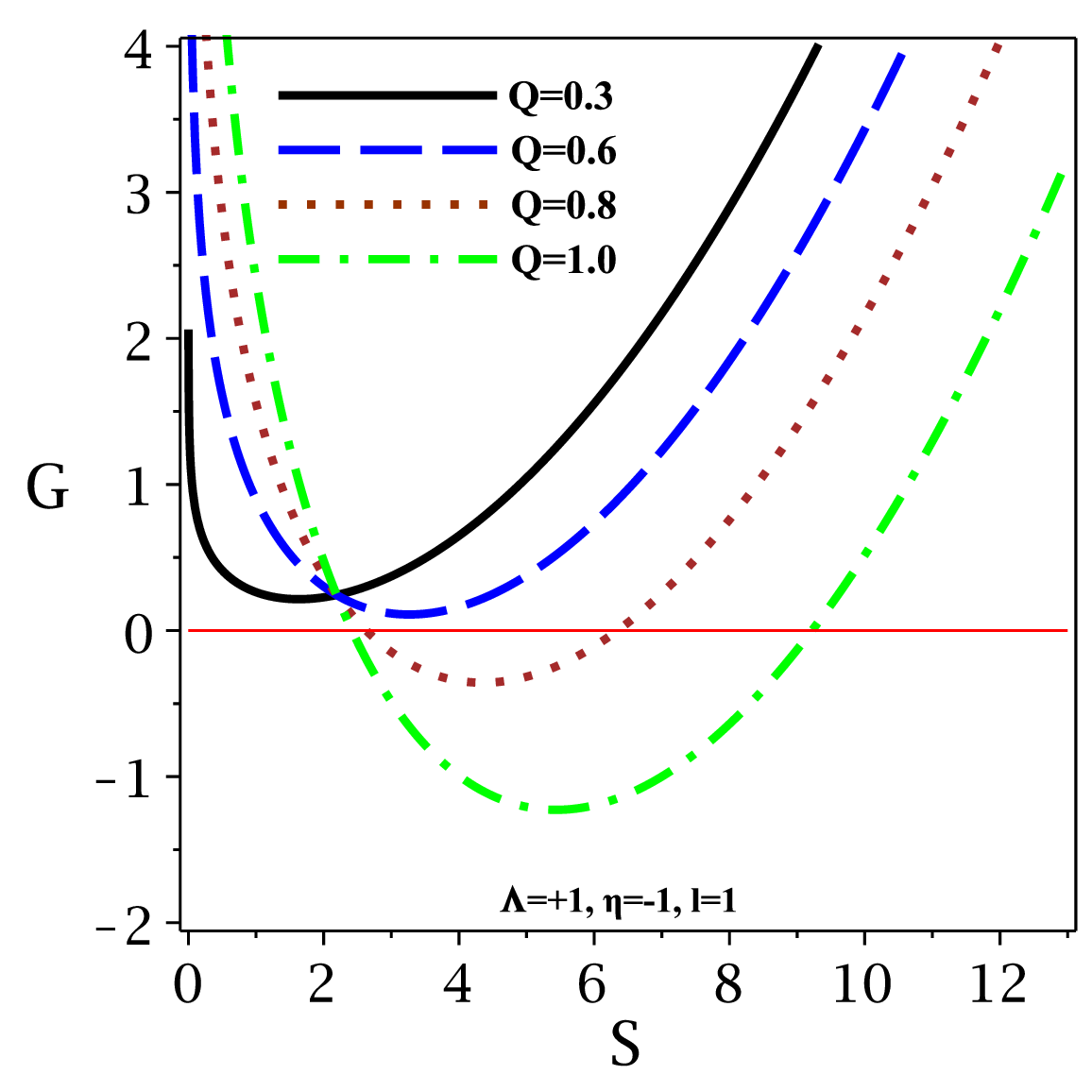} \newline
\includegraphics[width=0.35\linewidth]{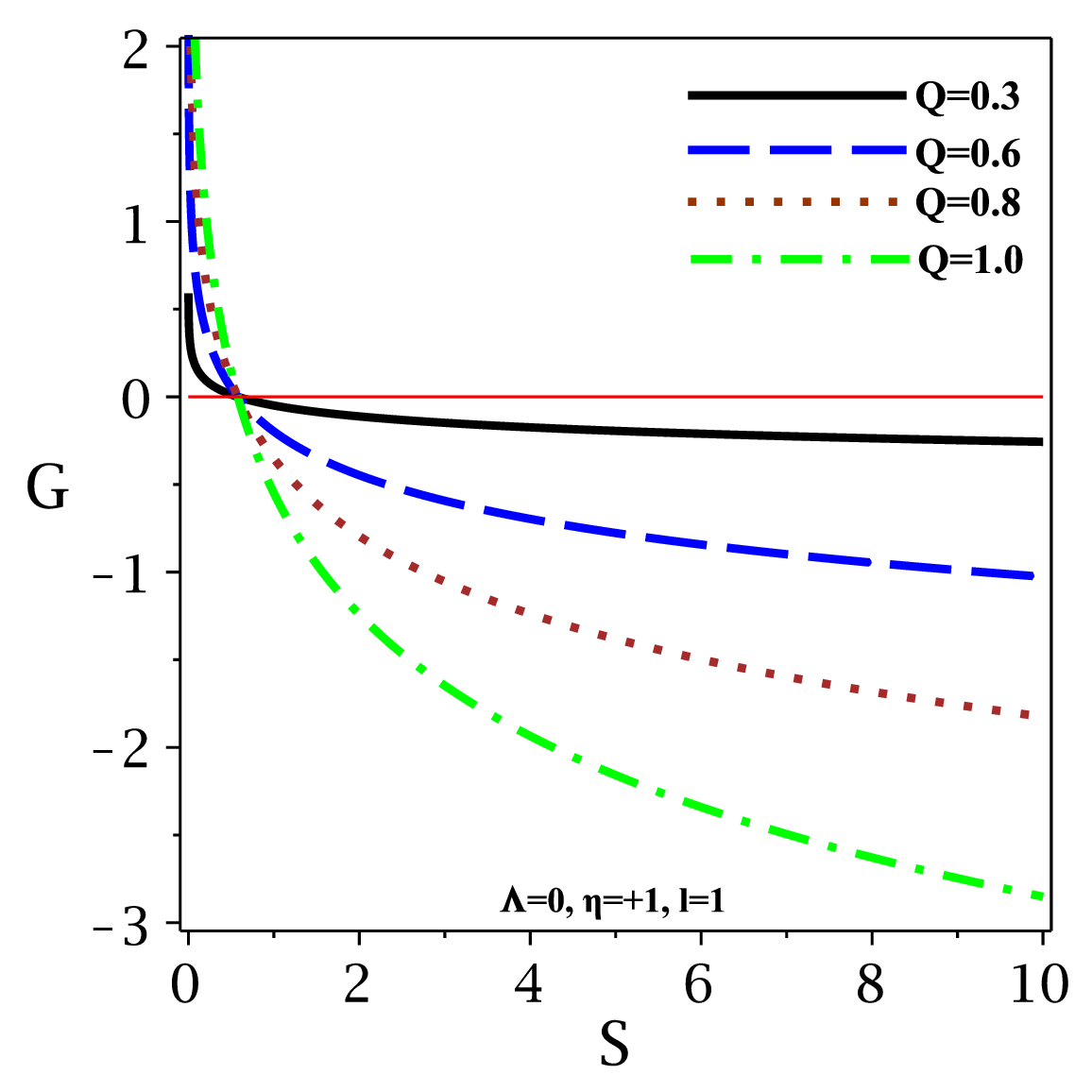} \includegraphics[width=0.35%
\linewidth]{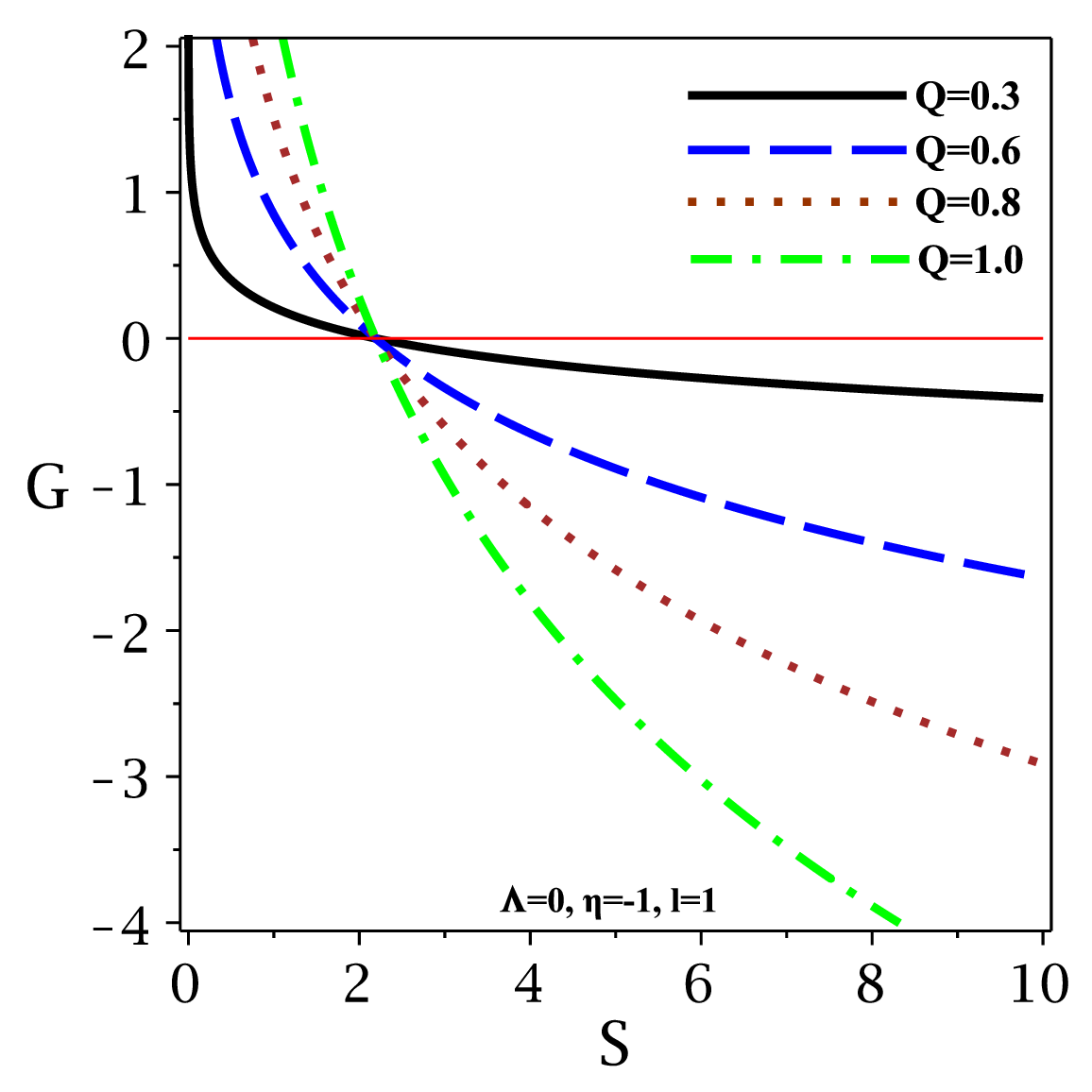} \newline
\caption{The Gibbs free energy $G$ versus $S$ for different values of the
charge. Left panel for phantom ($\protect\eta =1$), and right panel for
Maxwell case ($\protect\eta =-1$). Up panels for $\Lambda<0$, middle panels
for $\Lambda>0$, and down panels for $\Lambda=0$}
\label{Fig7}
\end{figure}

In order to gain a clear insight into the two conditions mentioned above, we plot the Gibbs free energy versus entropy in Fig. \ref{Fig7} to determine the areas of global stability. Our findings confirm that there is a global stability area for different values of the cosmological constant for BTZ black holes in the presence of both phantom and Maxwell fields. However, the global stability area is more extensive for phantom BTZ black holes compared to Maxwell's case. Furthermore, the global stability area changes with the variation of the electrical charge ($Q$). These changes are:

i) In the case of $\Lambda <0$ (AdS case); the global stability area decreases as the electrical charge increases (see the up panels in Fig. \ref{Fig7}).

ii) In the case of $\Lambda >0$ (dS case); there exists a critical value for the electrical charge ($Q_{critical}$). For $Q<Q_{critical}$, there is no global stability area, but for $Q>Q_{critical}$, this area appears and increases as the charge increases (see the middle panels in Fig. \ref{Fig7}).

iii) For $\Lambda =0$; large BTZ black holes satisfy the global stability condition. Furthermore, the global stability area is independent of the electrical charge (see the down panels in Fig. \ref{Fig7}).

\section{Conclusions}

In this work, the phantom BTZ black hole solutions were extracted in
three-dimensional spacetime for the first time. Then, the Ricci and
Kretschmann scalars of the obtained solutions were studied to find the
curvature singularity. Also, the asymptotical behavior of phantom BTZ black
holes was investigated. Our findings indicated that a curvature singularity
exists at $r=0$, and the asymptotical behavior could be (A)dS. To extract
different behaviors between the BTZ black holes in the presence of Maxwell
and phantom (anti-Maxwell) fields, the effects of different parameters
(mass, charged, and the cosmological constant) on the horizon of phantom BTZ black holes were studied. Indeed, phantom BTZ black holes with charged BTZ black holes were compared. Our findings indicated that there were three
substantially different behaviors of BTZ black holes between Maxwell and
phantom fields, which were:  i) There was only one root for phantom (anti-Maxwell) BTZ black holes. However, the roots of BTZ black holes changed in the presence of the Maxwell field. ii) The radius of the phantom BTZ black hole increased with the increase of electric charge, which was different from the Maxwell case. Moreover, the number of roots decreased with increasing electric charge. Our analysis indicated that BTZ black holes can have two roots (inner and outer roots), one root (extremal black holes), and no root (naked singularities) as the electric charge increases. iii) We found that phantom BTZ black holes can exist for $\lambda <0$, $\Lambda >0$ and $\lambda =0$. However, the BTZ black hole with Maxwell field only exists for the case $\Lambda <0$.

The study focused on obtaining the thermodynamic quantities of BTZ black holes with Maxwell and phantom fields. It was found that these black holes satisfied the first law of thermodynamics. Interestingly, the results showed that the obtained phantom BTZ black holes had a positive mass for all values of the cosmological constant ($\Lambda=0$, $\Lambda>0$, and $\Lambda<0$). Specifically, a constraint was identified in the form of $\Lambda < \frac{2\eta q^{2}}{r_{+}^{2}}\ln \left( \frac{r_{+}}{l}\right)$, to have a positive mass for the BTZ black holes. This constraint implies that the mass of the phantom BTZ black holes is always positive when $\Lambda < 0$ or $\Lambda = 0$. However, for $\Lambda > 0$, we must respect this constraint. On the other hand, the mass of the Maxwell BTZ black hole was found to be positive only for negative values of the cosmological constant.

The local stability of three-dimensional black holes has been studied in the context of the canonical ensemble using the heat capacity. Our analysis revealed two different behaviors for the heat capacity of the BTZ black hole in the presence of phantom and Maxwell fields. These are as follows: 

i) For $\Lambda>0$ (dS case), medium-sized phantom BTZ black holes were found to be stable. However, charged BTZ black holes in the presence of the Maxwell field were found to be unstable. 

ii) For $\Lambda<0$ (AdS case), large black holes in the presence of phantom and Maxwell fields exhibited local stability. However, there was a distinct stable region. In other words, phantom BTZ black holes had a larger stable area compared to the Maxwell case.

The global stability of BTZ black holes in the presence of phantom and Maxwell fields has been evaluated in the context of the grand canonical ensemble by calculating the Gibbs free energy. Our analysis indicated that the BTZ black holes can satisfy the global stability condition in the presence of phantom and Maxwell fields for all values of the cosmological constant ($\Lambda=0$, $\Lambda>0$, and $\Lambda<0$). However, the global stability area was more extensive for phantom BTZ black holes compared to Maxwell's case. In addition, we have studied the effect of electrical charge on the global stability area. The results revealed that: i) for $\Lambda <0$, the global stability area decreases as the electrical charge increases. ii) for $\Lambda >0$ there was a critical value for the electrical charge ($Q_{critical}$), below which there was no global stability area, but above which this area appeared and increased as the charge increased. iii) For $\Lambda =0$, large BTZ black holes satisfied the global stability condition. Furthermore, the global stability area was independent of the electrical charge.

In future work, we plan to study the stability of this solution using quasi-normal modes. We intend to generalize the solution and couple a canonical or phantom scalar field to the action, and see what contribution this coupling makes to thermodynamics.
 
\begin{acknowledgements}
B. Eslam Panah thanks University of Mazandaran..
\end{acknowledgements}

\end{document}